\documentclass[floats,twocolumn,aps,prb,longbibliography,superscriptaddress]{revtex4-2}

\usepackage{graphicx}
\usepackage{amsfonts}
\usepackage{amssymb}
\usepackage{dsfont}
\usepackage{amsbsy}
\usepackage{amsmath}\usepackage[colorlinks = true,
            linkcolor = magenta,
            urlcolor  = magenta,
            citecolor = magenta,
            anchorcolor = magenta]{hyperref}
\usepackage{hyperref}
\usepackage{xcolor}
\usepackage{subeqnarray}

\begin{document}

\title{Spin Liquids on the Tetratrillium Lattice}

\author{Mat\'ias G. Gonzalez}
\affiliation{Institute of Physics, University of Bonn, Nussallee 12, 53115 Bonn, Germany}
\affiliation{Helmholtz-Zentrum Berlin f\"ur Materialien und Energie, Hahn-Meitner-Platz 1, 14109 Berlin, Germany}
\affiliation{Dahlem Center for Complex Quantum Systems and Fachbereich Physik, Freie Universit\"at Berlin, Arnimallee 14, 14195 Berlin, Germany}

\author{Johannes Reuther}
\affiliation{Helmholtz-Zentrum Berlin f\"ur Materialien und Energie, Hahn-Meitner-Platz 1, 14109 Berlin, Germany}
\affiliation{Dahlem Center for Complex Quantum Systems and Fachbereich Physik, Freie Universit\"at Berlin, Arnimallee 14, 14195 Berlin, Germany}

\begin{abstract} 
The tetratrillium lattice has recently been proposed as responsible for the dynamical properties observed in the $S=1$ langbeinite compound K$_2$Ni$_2$(SO$_4$)$_3$. Here, we study in detail the classical spin liquid properties of this lattice of tri-coordinated tetrahedra using classical Monte Carlo and large-$N$ theory calculations. In the large-$N$ limit, we find that the system presents a gapped spectrum with flat bottom bands, giving rise to a fragile spin liquid with exponentially decaying correlations according to the classification of classical spin liquids. We confirm that this scenario also holds in the more realistic Ising and Heisenberg cases, for which the system does not exhibit any finite-temperature phase transition, and the low-temperature spin structure factors exhibit excellent quantitative agreement with the large-$N$ theory. We also provide insight into the quantum $S=1/2$ limit by performing pseudo-Majorana functional renormalization group calculations at finite temperatures, and discuss the possible phases that can arise in the ground state due to quantum fluctuations.
\end{abstract}

\date{\today}

\maketitle 

\section{Introduction}

Spin liquids (SLs) are states of matter that lack long-range magnetic order even in the absence of thermal fluctuations at zero temperature~\cite{Villain79, Moessner98, Balents10, Savary17, Broholm20}. In the \textit{classical} case, a CSL has a macroscopically (either extensively or subextensively) degenerate ground state, whose degeneracy is not lifted by entropic effects at finite temperature through order-by-disorder mechanisms~\cite{Villain79, Moessner98, Balents10}. When extensive, the degeneracy becomes evident by the presence of at least one flat band at the bottom of the spectrum~\cite{Davier23, Yan24a, Yan24b}. Very recently, a thorough classification of extensively degenerate CSLs has been developed based on the topology of the dispersive bands, which separates them into two groups~\cite{Yan24a, Yan24b}. On one hand, if the dispersive bands are gapless (with respect to the flat ones), the touching points define the type of emergent Gauss law that appears in the gauge theory (typically associated with pinch-point singularities in the spin structure factor). This gives rise to slow decaying correlations and a U(1) CSL. On the other hand, if the dispersive bands are gapped, the ground state exhibits exponentially decaying correlations, resulting in a fragile topological CSL.

Altogether, CSLs offer an interesting route to realizing exotic emergent phenomena, as the inclusion of quantum fluctuations \textit{may} lead to a quantum spin liquid (QSL) by creating a macroscopic superposition of states through tunneling between different classical ground states. A typical example of this process is classical spin ice (CSI) on the pyrochlore lattice Ising model~\cite{Bramwell20}. In this classical model, the ground states fulfill the two-in-two-out rule on each tetrahedron, and the system can fluctuate within this manifold by flipping spins on hexagonal plaquettes. On the other hand, flipping single spins breaks the ice rule on two tetrahedra and creates a pair of magnetic monopoles, which can then move freely through the lattice. The quantum counterpart, quantum spin ice (QSI), can be achieved by adding small perpendicular interactions to CSI, namely $S_i^+S_j^-$~\cite{Hermele04, Gingras14}. It has been shown by perturbation theory that these terms generate the necessary ring-exchange processes to turn the classical ground-state manifold into a quantum superposition leading to a QSL. QSI has received much attention in recent years due to a collection of interesting emergent phenomena, such as the presence of photon, vison, and fractional spinon excitations in the spectrum~\cite{Ross11, Benton12, Szabo19}. The interest in the community has been further boosted by the existence of possible material realizations such as Ho$_2$Ti$_2$O$_7$ and Dy$_2$Ti$_2$O$_7$ for CSI~\cite{Harris97, Ramirez99, Bramwell01, Lago07}, and the dipolar-octupolar compounds Ce$_2$Zr$_2$O$_7$ and Ce$_2$Sn$_2$O$_7$ for QSI~\cite{Gao19, Sibille20, Smith22, Bhardwaj22, Yahne24, Gao25}.

Another family of compounds that has received some attention lately in the context of searching for SL candidates is the langbeinite family~\cite{Zivkovic21, Boya22, Yao23, Gonzalez24, Khatua24, Kolay24, Kubickova24, Boya25, Sebastian25, Sebastian25b}. The members of this family have two symmetry-inequivalent sites forming two intertwined trillium lattices [see Fig.~\ref{fig:tetratrillium}(a) and (b)]. This structure provides several exchange paths with similar distances, which give rise to highly frustrated models, some of which have already been proposed as SL candidates, such as KSrFe$_2$(PO$_4$)$_3$~\cite{Boya22, Sebastian25} and Pb$_{1.5}$Fe$_2$(PO$_4$)$_3$~\cite{Boya25}. Many family members have very low ordering transition temperatures compared to the Curie-Weiss temperature that marks the onset of magnetic correlations. In particular, the $S=1$ langbeinite K$_2$Ni$_2$(SO$_4$)$_3$ has a Curie-Weiss temperature of $\sim -18$~K and presents a phase transition at $\sim 1$~K which only releases $\sim 1\%$ of the total entropy, resulting in very weak Bragg peaks accompanied by highly dynamical properties down to very low temperatures~\cite{Zivkovic21, Yao23}. Furthermore, K$_2$Ni$_2$(SO$_4$)$_3$ can be driven into a SL by introducing a small magnetic field~\cite{Zivkovic21}. Due to these properties, K$_2$Ni$_2$(SO$_4$)$_3$ has been proposed to lie close to a SL region centered around a previously unexplored lattice consisting of tri-coordinated tetrahedra (in contrast to the pyrochlore lattice, which is bipartite in terms of tetrahedra) dubbed \textit{tetratrillium} lattice~\cite{Gonzalez24}. 

In this context, here we study in detail the spin liquid properties of the tetratrillium lattice in several classical limits using classical Monte Carlo simulations for the Ising and Heisenberg models and large-$N$ theory for the $N\to \infty$ limit (where $N$ is the number of components of the spin vector). In the large-$N$ limit, we find a gap that separates the bottom flat bands from the dispersive ones. According to the SL classification scheme~\cite{Yan24a, Yan24b}, the system hosts a fragile topological SL with exponentially decaying correlations. We verify this by calculating the spin structure factor along several planes, which does not exhibit any ordering peaks or pinch-point singularities (associated with gapless classical SLs). We confirm this picture for the Heisenberg and Ising cases, where we find no evidence of a finite-temperature phase transition, and the spin structure factors show an excellent agreement compared to the large-$N$ theory. For the Ising case, we use previous knowledge from the trillium lattice to calculate an estimate of the remaining entropy at zero temperature and develop a cluster-update algorithm for the tetratrillium lattice. Finally, we perform pseudo-Majorana functional renormalization group calculations to study the quantum $S=1/2$ Heisenberg limit at finite temperatures. In this case, we also do not observe any signatures of a finite-temperature phase transition down to the lowest reachable temperatures, and obtain a spin structure factor without clear peaks and with a more diffuse structure resembling the classical one but broadened by quantum fluctuations.

The remainder of the article is organized as follows. In Section~\ref{sec:modandmet}, we present the tetratrillium lattice and the methods we use to obtain our results: large-$N$ theory, classical Monte Carlo, and pseudo-Majorana functional renormalization group. Then we present our main results in Sec.~\ref{sec:res}, first for the classical case in the $N\to\infty$ limit and then for the Ising and Heisenberg cases, and finally for the quantum $S=1/2$ Heisenberg model. Last, we present our concluding remarks in Sec.~\ref{sec:conc}.

\section{Model and Methods}
\label{sec:modandmet}

\begin{figure}[t!]
    \centering
    \includegraphics[width=0.95\linewidth]{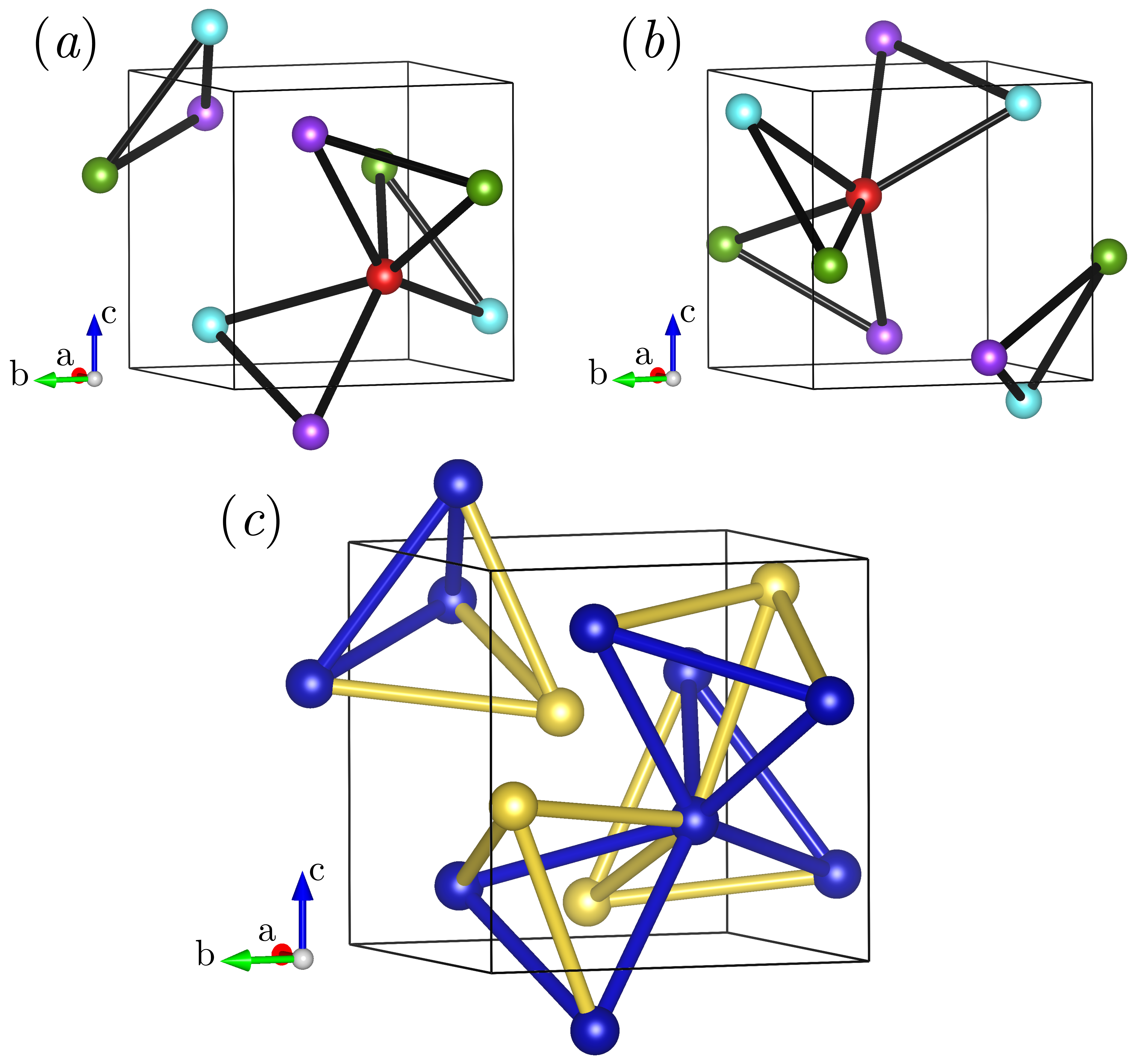}
    \caption{(a) Trillium lattice unit cell for $u=0.33554$ and  (b) $u=0.59454$, where the 4 triangles in the unit cell are shown. Each site in the unit cell is shown in a different color. (c) Tetratrillium lattice unit cell using the same $u$ values [blue for (a) sites and yellow for (b) sites]. Only the 4 tetrahedra in the unit cell are displayed.}
    \label{fig:tetratrillium}
\end{figure}

The tetratrillium lattice can be built from two symmetry-inequivalent trillium lattices, so let us start by describing them individually. The trillium lattice has a simple cubic unit cell with 4 sites, each belonging to 3 triangles. Therefore, there are 4 triangles per unit cell or, equivalently, one triangle per site. The positions of the 4 spins in the cubic unit cell of side length $a=1$ can be written according to Table~\ref{tab:trillium1}, where $u$ is a free parameter~\cite{Hopkinson06, Isakov08}. In Fig.~\ref{fig:tetratrillium} we show two examples for $u=0.33554$ [panel (a)] and $u=0.59454$ [panel (b)], which are the values corresponding to K$_2$Ni$_2$(SO$_4$)$_3$ at low temperatures~\cite{Gonzalez24}. Throughout the article, we will use these values to calculate position-dependent quantities such as the spin structure factor. For the two trillium lattices, the sites 1, 2, 3, and 4 from Table~\ref{tab:trillium1} are displayed in red, green, light-blue, and purple in Fig.~\ref{fig:tetratrillium}(a) and (b), respectively. The bonds forming each trillium lattice are shown in black, and only the 4 triangles in the unit cell are depicted. Let us define from now on that one trillium lattice has nearest-neighbor coupling $J_1$ and the second one has nearest-neighbor coupling $J_2$.

\begin{table}[h!]
    \begin{tabular}{ccc}
    site &\ & position \\ \hline \hline
    1 &\ &$\left(u,u,u \right)$ \\
    2 &\ &$\left(-\frac{1}{2}+u,\frac{1}{2}-u,1-u \right)$  \\
    3 &\ &$\left(1-u,-\frac{1}{2}+u,\frac{1}{2}-u \right)$  \\
    4 &\ &$\left(\frac{1}{2}-u,1-u,-\frac{1}{2}+u \right)$ 
    \end{tabular}
    \caption{Positions of the four sites corresponding to a trillium lattice within the cubic unit cell.}
    \label{tab:trillium1}
\end{table}

Then, to build the tetratrillium lattice, each site of the second trillium lattice is connected via $J_{12}$ to the closest $J_1$ triangle of the first trillium lattice, in such a way that it forms a tetrahedron. Throughout the article, we take $J_1=J_{12}=1$ and $J_2=0$ to form the nearest-neighbor tetratrillium lattice. In Fig.~\ref{fig:tetratrillium}(c), lattice sites from the $u=0.33554$ trillium lattice are connected ($J_1$) and shown in blue, while the sites from the other trillium lattice are shown in yellow and are disconnected from one another ($J_2=0$). Then, each yellow site is connected to 3 blue ones, forming tetrahedra ($J_{12}=J_1$). The 4 tetrahedra shown in Fig.~\ref{fig:tetratrillium}(c) are the ones in the unit cell, which are repeated and translated throughout the lattice. In total, there are 8 sites per unit cell divided into two symmetry-inequivalent groups (corresponding to each of the trillium lattices). Each of the sites from the $J_1$ trillium lattice belongs to 3 tetrahedra. Each of the sites from the $J_2$ trillium lattice belongs to 1 tetrahedron. Therefore, there are 2 sites per tetrahedron, which means that there are 4 tetrahedra per unit cell (same as the number of triangles in the trillium lattice). 

Even though there is no exact realization of the tetratrillium lattice in a real material, there exists the $S=1$ langbeinite compound K$_2$Ni$_2$(SO$_4$)$_3$ which lies rather close in parameter space~\cite{Zivkovic21, Yao23, Gonzalez24}. In our current terminology of couplings, the exchange parameters derived in Ref.~\cite{Gonzalez24} using density-functional theory energy mappings correspond to $J_1=0.48~J_{12}$, and $J_2=0.14~J_{12}$, with other inter-trillium couplings less than $7\%$ of $J_{12}$. On the other hand, the authors in Ref.~\cite{Yao23} used self-consistent Gaussian approximation to obtain $J_1=0.55~J_{12}$, and $J_2=0.02~J_{12}$, with other inter-trillium couplings less than $6\%$ of $J_{12}$. Overall, both approaches predict that $J_1$ and $J_{12}$ are the most important couplings, where $J_{1}$ is approximately half of $J_{12}$. This means that K$_2$Ni$_2$(SO$_4$)$_3$ lies in proximity to the spatially-isotropic tetratrillium lattice explained above.

To gain insights into the physics of the tetratrillium lattice, we use large-$N$ theory and classical Monte Carlo (cMC) calculations to solve the classical Ising ($N=1$), Heisenberg ($N=3$), and $N=\infty$ cases. For the quantum $S=1/2$ case, we use the pseudo-Majorana functional renormalization group (PMFRG) method to compute the correlations at finite temperatures. All these methods are explained below.

\subsection{Large-$N$ theory}

A spin Hamiltonian on a lattice can be generically written as 
\begin{equation}
    \mathcal{H} = \frac{1}{2} \sum_{i,j}^M \sum_{\alpha,\beta}^{N_\text{sub}} J_{{i_\alpha},{j_\beta}} \mathbf{S}_{i_\alpha} \mathbf{S}_{j_\beta},
\label{eq:ham1}
\end{equation}
where $i$ and $j$ run over a Bravais lattice containing $M$ unit cells, and $\alpha$ and $\beta$ run over the $N_\text{sub}$ sites in the unit cell. The factor $1/2$ corrects the double counting and $\mathbf{J}$ is the $(MN_\text{sub}) \times (MN_\text{sub})$ interaction matrix with elements $J_{{i_\alpha},{j_\beta}}$. Finally, $\mathbf{S}_{i_\alpha}$ are the classical spin vectors of $N$ components. 

In the case of the tetratrillium lattice, and considering that we have set $J=1$, we can rewrite the Hamiltonian in the \textit{constrainer form} as follows
\begin{equation}\label{eq:constrainer_ham}
    \mathcal{H} \equiv \sum_{i=1}^{M} \sum_{\boxtimes=1}^4 T_{i,\boxtimes}^2.
\end{equation}
where $\boxtimes$ runs over each of the four tetrahedra in the unit cell (see Fig.~\ref{fig:tetratrillium}). Then, $T_{i,\boxtimes}^2 = (S_{i,\boxtimes_1} + S_{i,\boxtimes_2} + S_{i,\boxtimes_3} + S_{i,\boxtimes_4})^2$ is the squared sum of the four spins in each tetrahedron $\boxtimes$. This Hamiltonian is equivalent to the one in Eq.~\eqref{eq:ham1} but has an energy shift related to the extra $S_{i,\boxtimes_j}^2=1$ constant terms. Since the tetratrillium lattice has 4 constraints and 8 sites per unit cell, there must be 4 bottom flat bands and 4 dispersive upper bands~\cite{Yan24a, Yan24b}.

Using the constrainer form Hamiltonian and following Ref.~\cite{Yan24b}, one can easily obtain the Fourier-transformed coupling matrix $\mathbf{J}(\mathbf{q})$, which is the $N_\text{sub} \times N_\text{sub}$ matrix with elements $J_{\alpha,\beta}(\mathbf{q})$. Its eigenvalues give direct access to the spectrum and the $T=0$ spin correlations. Here, it is important to note an advantage of using the constrainer-form Hamiltonian. When using the Hamiltonian in Eq.~\eqref{eq:ham1}, and since there are two symmetry-inequivalent sites, two different energy shifts for the two sublattices have to be added to the diagonals of $\mathbf{J}(\mathbf{q})$, to ensure that the spins on the two sublattices obey the same normalization [which is indicated by the components of the eigenvectors of $\mathbf{J}(\mathbf{q})$]. However, the constrainer form naturally accounts for this by setting the minimum energy to zero with energy shifts that depend on the number of bonds as $N_\textrm{bonds}S_i^2$ (which is different for spins in the two sublattices). 

At finite temperatures, the constraint on the spin length is addressed by a Lagrange multiplier $\lambda_\alpha$, which is translation invariant but, in principle, different on each site of the unit cell. The partition function then results in
\begin{equation}
    \mathcal{Z}(T) \propto \int D\lambda \ e^{\frac{N}{2} \text{Tr}\left[ \sum_\mathbf{q} \log \mathbf{A} - M \boldsymbol{\lambda}\right]},
\end{equation}
where $\mathbf{A} = \beta \mathbf{J}(\mathbf{q})+\boldsymbol{\lambda}$, $\beta$ is the inverse temperature, and $\boldsymbol{\lambda}$ is the diagonal matrix that contains the values of $\lambda_\alpha$.

The Taylor expansion up to quadratic order becomes exact in the $N\to\infty$ limit~\cite{Berlin52, Stanley68, Moshe03}, and thus $\mathcal{Z}(T)$ can be integrated exactly. The minimization with respect to $\boldsymbol{\lambda}$ gives the saddle-point condition,
\begin{equation}
    \frac{1}{M} \sum_\mathbf{q} \left[ \left( \beta \mathbf{J}(\mathbf{q})+\boldsymbol{\lambda} \right)^{-1} \right]_{\alpha,\alpha} = 1,
    \label{eq:self}
\end{equation}
where the subindex of the bracket indicates the element of the inverse of $\beta \mathbf{J}(\mathbf{q})+\boldsymbol{\lambda}$. For symmetry equivalent sites, $\lambda_\alpha$ is the same, and the calculations can be simplified. In our case, we have two types of sites corresponding to each trillium lattice. Thus, we have two different values $\lambda_1$ and $\lambda_2$ repeated four times each on the diagonal matrix $\boldsymbol{\lambda}$. For a given temperature, we then solve two coupled equations iteratively by choosing a starting point $\{ \lambda_1^0,\lambda_2^0\}$ and updating one value at a time using the corresponding equation, resulting in the series
\begin{equation}
    \{ \lambda_1^0,\lambda_2^0\} \to \{ \lambda_1^1,\lambda_2^0\} \to \{ \lambda_1^1,\lambda_2^1\} \to ... \to \{ \lambda_1^n,\lambda_2^n\}.
    \label{eq:self2}
\end{equation}
Once converged parameters $\lambda_\alpha$ are obtained, the spin structure factor at finite temperatures can be calculated as
\begin{equation}
    S(\mathbf{q}) = \sum_{\alpha, \beta} \left[ \beta \mathbf{J}(\mathbf{q}) + \boldsymbol{\lambda}\right]^{-1}_{\alpha,\beta}.
    \label{eq:ssflargen}
\end{equation}

\subsection{Classical Monte Carlo}

As mentioned in the previous subsection, the large-$N$ theory is exact in the unphysical limit where the spin has an infinite number of components. The theory is useful for gaining an initial insight into the system's properties. For example, a recent classification of classical spin liquids has been developed using the band structures in this limit~\cite{Yan24a, Yan24b}. However, it is not guaranteed that a spin liquid in the $N\to \infty$ limit will survive down to the more realistic Heisenberg ($N=3$) and Ising ($N=1$) cases. Two examples are the kagome Heisenberg model, which presents an order-by-disorder transition~\cite{Chalker92, Zhitomirsky08, Chern13}, and the trillium lattice, which has a unique ordered ground state in the Heisenberg case~\cite{Isakov08}. Therefore, it is imperative to perform classical Monte Carlo (cMC) simulations to check for the validity of the large-$N$ solutions.

For the Ising case, we treat the spins as discrete variables that can take on the values $-1$ and $1$. Then we start at high temperatures from a random spin configuration and perform single-spin update trials. Each Monte Carlo step consists of $MN_\text{sub}$ trials, and we perform $5\times10^5$ steps at each temperature. We use the first half of the steps for thermalizing and the second half for measuring thermodynamic quantities. We use a logarithmic cooling protocol from $T=100~J$ down to $0.01~J$ with 200 steps. We consider cubic systems with $N_s = N_\text{sub} L^3$ sites, with $L$ ranging from 2 to 15 (27000 spins). We also measure correlations between spins to calculate the spin structure factor by using previously thermalized configurations at specific temperatures. The cMC simulations on these kinds of Ising systems usually have the problem that single-spin updates become unlikely at low temperatures (because flipping a spin costs too much energy), and, as a result, the system does not explore the ground-state manifold properly (or in acceptable computing times). To overcome this problem, we use the usual method of implementing loop updates. We explain these details in Sec.~\ref{sec:res}, where we discuss properties of the ground state manifold. 

In the Heisenberg case, the spins are three-dimensional unit vectors. The difference to the Ising case is that the spins can be updated continuously. This is done by using the adaptive Gaussian step, which guarantees a $50\%$ acceptance ratio at all temperatures~\cite{Alzate19}. In addition, for each spin update trial, we perform two overrelaxation steps to improve thermalization. In both the Ising and Heisenberg cases, we perform 10 independent runs to improve statistics and obtain smoother results.

\subsection{Pseudo-Majorana Functional Renormalization Group}

The pseudo-Majorana functional renormalization group (PMFRG) method relies on rewriting the quantum $S=1/2$ spin operators in terms of Majorana operators
\begin{equation}
S_i^x = -i\eta_i^y \eta_i^z ,\ \ S_i^y = -i\eta_i^z \eta_i^x ,\ \ S_i^z = -i\eta_i^x \eta_i^y
\end{equation}
that fulfill $\{\eta_i^\mu, \eta_j^\nu \}=\delta_{i,j} \delta_{\mu,\nu}$. This representation, in contrast to the fermionic one (PFFRG)~\cite{Reuther10, Muller24}, has the advantage that it does not introduce any unphysical states. Instead, it introduces trivial redundancy in the Hilbert space, which can be easily addressed~\cite{Niggemann21}. As a consequence, PMFRG can be applied at finite temperatures and becomes asymptotically exact in the $T\to \infty$ limit. The general FRG scheme typically works by introducing an infrared cutoff frequency $\Lambda$ which makes the system exactly solvable in the $\Lambda \to \infty$ limit, reaching the physical result in the $\Lambda \to 0$ limit. In the case of PMFRG, the $T$-flow has been recently developed, where the physical temperature is used as the flowing parameter $\Lambda$~\cite{Schneider24}. Within the PMFRG framework, spin correlations are determined from fermionic vertex functions that result from solving a hierarchy of coupled differential equations. The method relies on an approximation that introduces a cutoff, truncating this hierarchy at a specified order of the vertex functions. The PMFRG enables the calculation of finite-temperature phase transitions with remarkable accuracy through finite-size scaling analysis~\cite{Niggemann22, Schneider24, Schaden25}, as long as the transition temperature is not too low. Finally, the PMFRG (like the PFFRG) method preserves all (lattice and spin) symmetries, so nematic phases cannot be detected without inducing small perturbations and measuring their response~\cite{Hering22, Hagymasi22, Schaden25}. For further details on the method, we refer the interested reader to the more specialized literature~\cite{Niggemann21, Niggemann22, Schneider24}. 

Here we use the \texttt{TemperatureFlow} branch from the \texttt{PMFRG.jl} package~\cite{NiggemannGithub} to solve the quantum $S=1/2$ Heisenberg model on the tetratrillium lattice starting from the $T\to \infty$ limit and flowing down to low temperatures $T=0.01~J$. Throughout the flow, we calculate the spin correlations to obtain the equal-time spin structure factor and evaluate the magnetic susceptibility. As mentioned above, we assume that all lattice symmetries are preserved in our PMFRG calculations. Thus, we only need to calculate the RG temperature flow for the symmetry-inequivalent two-point correlators stemming from the two symmetry-inequivalent sites. The remaining ones in the unit cell are related to these by lattice rotations and reflections, and the ones outside the unit cell are related by lattice translations. 

\section{Results}
\label{sec:res}

In this section, we present the results for the tetratrillium lattice on different models. First, we discuss the classical models for the $N\to \infty$, Ising $(N=1)$, and Heisenberg $(N=3)$ cases. After that, we discuss the quantum $S=1/2$ Heisenberg case.

\subsection{The $N\to \infty$ limit}

As mentioned above, a classification scheme has been recently developed, which predicts the type of spin liquid in the large-$N$ limit~\cite{Yan24a, Yan24b}. In this limit, the spin components become independent, and the spins can therefore be treated as scalars. The classification is based on the structure of the bottom flat bands and the dispersive bands above them. When there is no gap between these, the ground state is an algebraic classical spin liquid, where the term algebraic refers to the decay of the correlations as a function of distance. On the other hand, if there is a gap between the flat bands and the other bands, the ground state is a fragile topological classical spin liquid, and the correlations decay exponentially.

\begin{figure}[t!]
    \centering
    \includegraphics[width=0.95\linewidth]{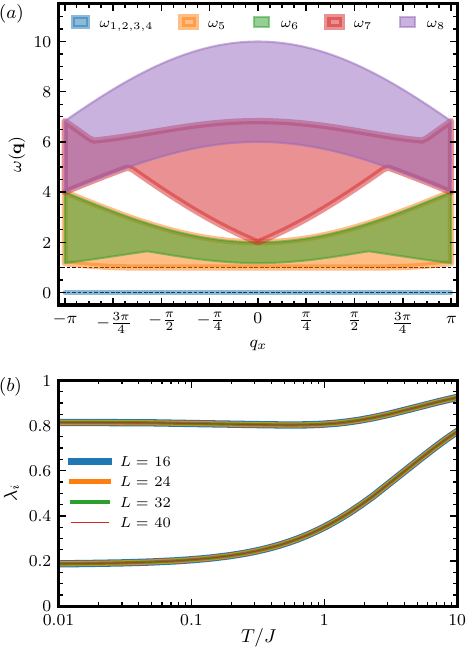}
    \caption{(a) Spectrum $\omega(\mathbf{q})$ of the coupling matrix $\mathbf{J}(\mathbf{q})$ in the first Brillouin zone at $T=0$, composed of 8 bands with a gap denoted by the dashed black lines. There are 4 degenerate flat bands at the bottom (blue). (b) Values of $\lambda_i$ as a function of the temperature for the two symmetry-inequivalent sublattices and different system sizes.}
    \label{fig:largen}
\end{figure}

Here, we diagonalize the coupling matrix $\mathbf{J}(\mathbf{q})$ obtained from the constrainer-form Hamiltonian in Eq.~(\ref{eq:constrainer_ham}), where the eigenvalues correspond to the spectrum $\omega(\mathbf{q})$. The results are shown in Fig.~\ref{fig:largen}(a), where we plot the three-dimensional bands as a contour that depends only on $q_x$. This is done by projecting $\omega(q_x,q_y,q_z) \to \omega(q_x)$ for all $q_y$ and $q_z$ in the first Brillouin zone and then extracting the profile of the lower and upper edge of each band. This allows us to see the existence of a gap of size $\Delta = 1$ (denoted by black dashed lines) between the four-fold degenerate flat bands at $\omega=0$ (in blue) and the dispersive bands (all other colors). According to Refs.~\cite{Yan24a, Yan24b}, a gapped spectrum implies a fragile topological classical spin liquid with exponentially decaying correlations down to zero temperature. This family of spin liquids is the classical analog of $\mathbb{Z}_2$ quantum spin liquids, with only a few known examples: the two-dimensional ruby and kagome lattices with additional couplings~\cite{Rehn17} and the three-dimensional trilline lattice~\cite{Fancelli25}. These systems are in contrast to the more common gapless U(1) spin liquids, which typically arise in bipartite structures of corner-sharing clusters, and give rise to emergent Gauss laws and pinch-point singularities in the spin structure factor.

\begin{figure*}[t!]
    \centering
    \includegraphics[width=0.9\linewidth]{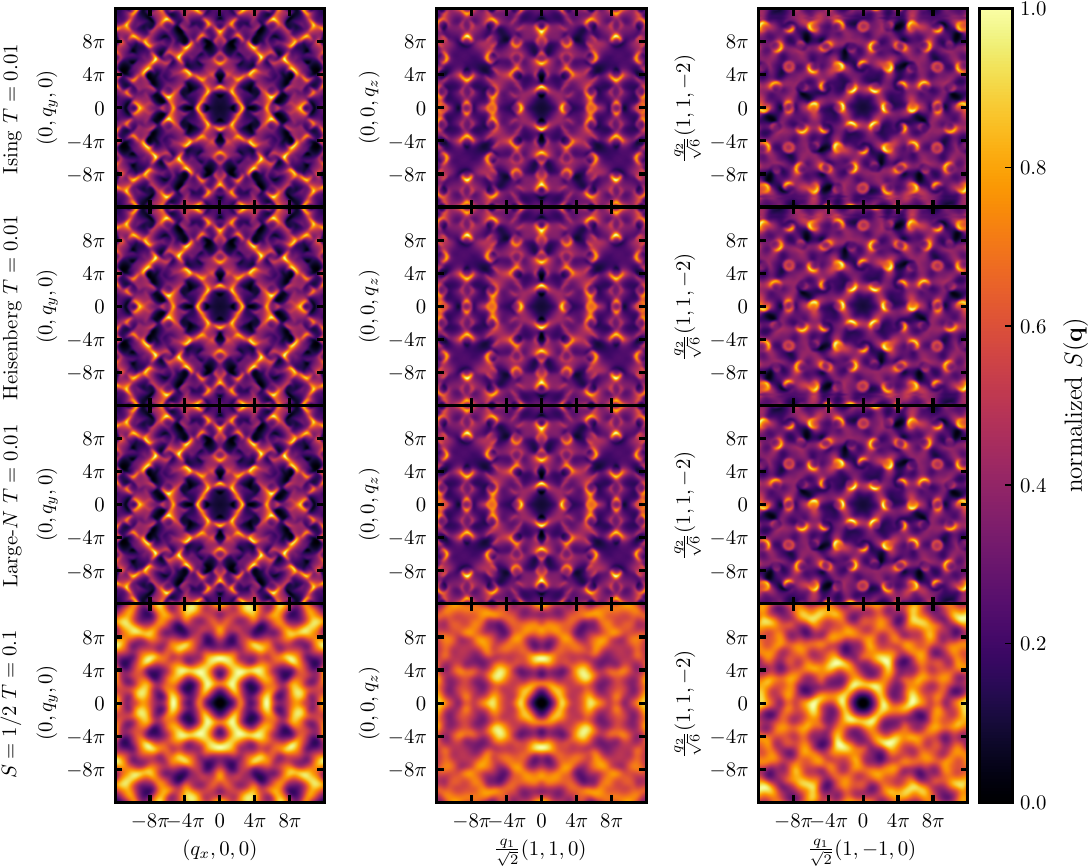}
    \caption{Spin structure factor on the tetratrillium lattice for three different planes in the Brillouin zone, displayed one in each column. Each row represents a different model, where the first three correspond to the classical Ising, Heisenberg, and Large-$N$ models at low temperatures, $T=0.01$. The last row corresponds to the PMFRG results for the quantum $S=1/2$ Heisenberg model at $T=0.1$. All panels are normalized individually to their respective maximums.}
    \label{fig:ssfall}
\end{figure*}

To further confirm the scenario of a fragile classical spin liquid on the tetratrillium lattice, we investigate the spin structure factor using Eq.~\eqref{eq:ssflargen}, which requires calculating $\lambda_i$ for the sites in the unit cell. Since there are two symmetry-inequivalent sites in the tetratrillium lattice, there are two different values of $\lambda$, one for each underlying trillium lattice. In the $T\to \infty$ limit, $\lambda_i=1$ for all sites. We start from that limit and solve the self-consistency equations while lowering the temperature. The results are shown in Fig.~\ref{fig:largen}(b). With these values, Eq.~\eqref{eq:ssflargen} can be used to calculate the spin structure factor. In the third row of Fig.~\ref{fig:ssfall}, we show the spin structure factor at $T=0.01$ for three different planes in the Brillouin zone: the $(q_x,q_y,0)$ plane in the first row, the $(q_x,q_x,q_z)$ plane in the second, and the $[111]$ plane in the last row. These are the same planes studied in Ref.~\cite{Gonzalez24}, where experimental results for the $S=1$ langbeinite K$_2$Ni$_2$(SO$_4$)$_3$ are presented. From Fig.~\ref{fig:ssfall}, it is clear that there are no diverging peaks in the spin structure factor, indicating the spin liquid nature of the model. Another important property is the absence of any pinch-point singularities in the spin structure factor, as expected for a fragile spin liquid.

It is constructive to compare these results with those for the single trillium lattice, where $N_\mathrm{sub}=4$ and there are 4 constraints per unit cell. Therefore, we do not expect any flat bands (which means that there are no extensively degenerate ground states). In fact, large-$N$ studies found only subextensively degenerate ground states arising from a degenerate surface in momentum space $\cos^2(q_x/2)+\cos^2(q_y/2)+\cos^2(q_z/2) = 9/4$ instead of flat bands~\cite{Hopkinson06}. This illustrates that adding extra spins to form tetrahedra from triangles preserves the number of constraints while increasing the number of sites per unit cell, $N_\mathrm{sub}$, thereby leading to the emergence of flat bands at the bottom of the spectrum. In our specific case, this turns a subextensively degenerate spiral spin liquid on the trillium lattice into a fragile, extensively degenerate spin liquid on the tetratrillium lattice.

\subsection{The Ising limit $(N=1)$}

We now turn to the opposite side of the $N$ range: the Ising model where $S_i=\pm1$. We will see in what follows that the system has an extensive ground-state degeneracy, accompanied by a finite entropy at $T=0$. Furthermore, we find that these properties are fundamentally linked to equivalent behaviors on the trillium lattice, which were already addressed in Ref.~\cite{Radpath10}.

\subsubsection{Number of ground states}

For the trillium lattice, the constraint $T_{i,\triangle}^2 = (S_{i,\triangle_1} + S_{i,\triangle_2} + S_{i,\triangle_3})^2$ never vanishes because three Ising spins can never cancel each other. Thus, the ground states are composed of 2-up-1-down or 1-up-2-down states, such that $T_{i,\triangle}^2 = 1$, for all triangles in the lattice. This means that only two out of the eight possible states in each triangle are excited states (the ones corresponding to all up or all down spins). In other words, out of eight possible states, six are ground-state building blocks. On the other hand, in the tetratrillium lattice, there are 16 possible states in each tetrahedron, resulting from the eight states in the triangle plus an extra up/down spin. Triangles with all up or down spins cannot be fixed by the extra spin (the sum of spins in the tetrahedron is still non-zero). On the other hand, 2-up-1-down or 1-up-2-down states in the triangles can be fixed by the extra spin such that $T_{i,\boxtimes}^2 = 0$, which results in six ground states out of the possible 16 states. This argument also implies that the ground-state configurations of the trillium and tetratrillium lattices are in one-to-one correspondence.

A rough estimate of the entropy can be obtained using Pauling's approximation of uncorrelated units~\cite{Pauling33} (triangles in the case of the trillium lattice and tetrahedra in the case of the tetratrillium lattice). For the trillium lattice, the residual entropy per site is $s= \frac{1}{N_s} \ln \left(2^{N_s} \Omega^{N_\triangle}\right)$, where $\Omega$ is the ratio of ground states to the total number of states in each unit. In this case, $\Omega = 6/8$ and $N_\triangle = N_s$ yielding $s= \ln \frac{3}{2}$. In the tetratrillium case, we have $\Omega = 6/16$ but $N_\boxtimes = N_s/2$. Therefore, in the tetratrillium lattice, the residual entropy is
\begin{equation}
    s = \frac{1}{N_s} \ln \left(2^{N_s} \Omega^{N_\boxtimes}\right) = \frac{1}{N_s} \ln \left( 4^\frac{N_s}{2} \frac{6}{16}^\frac{N_s}{2}\right) = \frac{1}{2} \ln \frac{3}{2},
\end{equation}
which is half of the trillium lattice case. 

Improved estimates can be obtained by explicitly counting the number of ground states on larger spin clusters. Specifically, for a system with $N_s=N_\text{sub}L^3$ sites and periodic boundary conditions, we can count the number of ground-state configurations $n$ and calculate the entropy for this system size directly as
\begin{equation}
    s = \frac{1}{N_s} \ln n.
\end{equation}
For $L=1$, we find $n=6$ ground states, and for $L=2$, there are $n=314874$ ground-state configurations, in agreement with Ref.~\cite{Radpath10}. The resulting entropies are presented in Table~\ref{tab:entropies}, where the tetratrillium value is always half of the trillium one. We can see in Table~\ref{tab:entropies} that for $L=2$ the result is very close to the naive Pauling estimate in the first row.

\begin{table}[t!]
    \begin{tabular}{c|c|cc}
        &       &  $s(T=0)$ \\
    unit & $n$ & trillium & tetratrillium \\ \hline
    $\triangle$ or $\boxtimes$ & 6 & 0.4055 & 0.2027 \\
    $L=1$ & 6 & 0.4479 & 0.2240 \\
    $L=2$ & 314874 & 0.3956 & 0.1978
    \end{tabular}
    \caption{Pauling's residual entropy at $T=0$ for the trillium and tetratrillium lattices, taking into account different units for the counting argument.}
    \label{tab:entropies}
\end{table}

\subsubsection{Elementary spin fluctuations and Hilbert space sectors}

\begin{figure}[!h]
    \centering
    \includegraphics[width=0.82
    \linewidth]{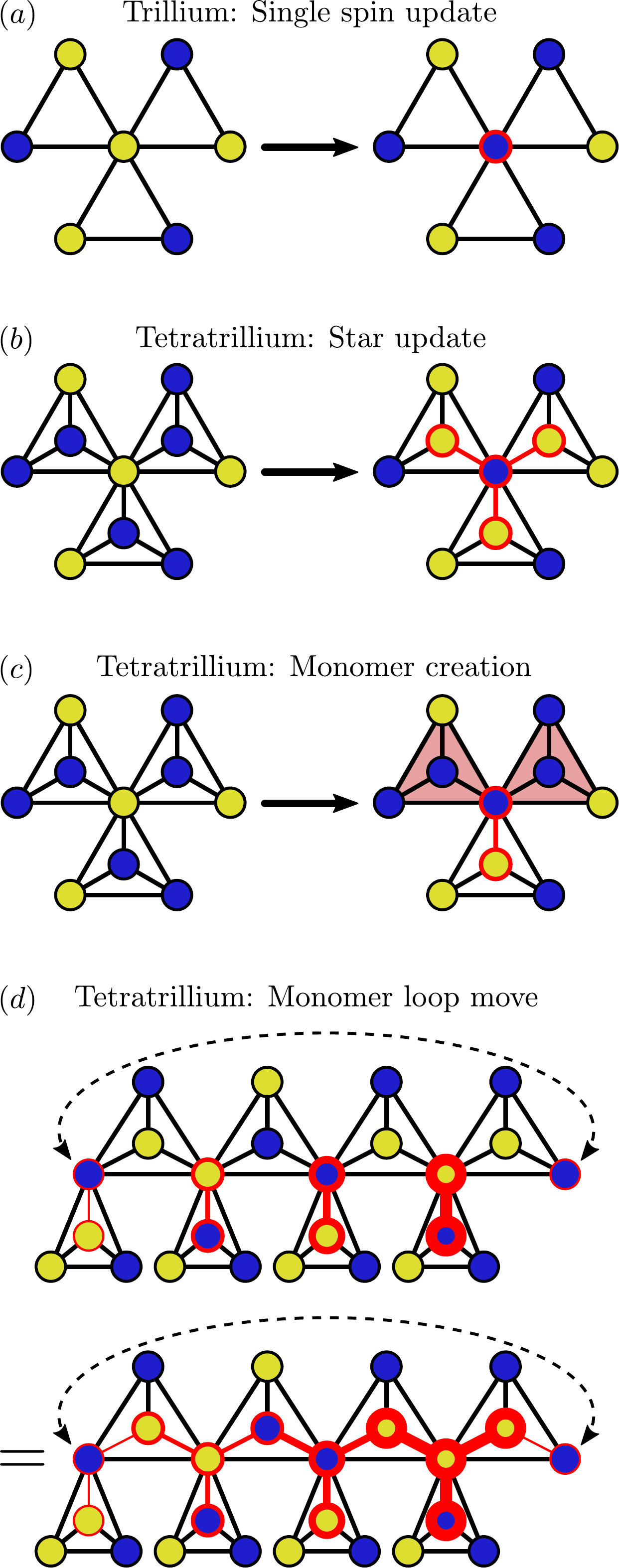}
    \caption{(a) Constant energy updates on the trillium lattice and (b) how they translate to star updates on the tetratrillium lattice. (c) Two monomers in the tetratrillium lattice (light red tetrahedra) arise from flipping a bond connecting the two trillium lattices. Blue and yellow sites indicate up and down spins, and the red circles indicate the spins that have been flipped. (d) A loop move of monomers in the top panel is created by flipping the red bonds in the depicted state in the order of increasing thickness (from left to right). The arrows indicate identical sites due to periodic boundaries. This process is identical to a sequence of star updates in the bottom panel (from left to right).}
    \label{fig:fliploop}
\end{figure}

\begin{figure}[t!]
    \centering
    \includegraphics[width=0.9\linewidth]{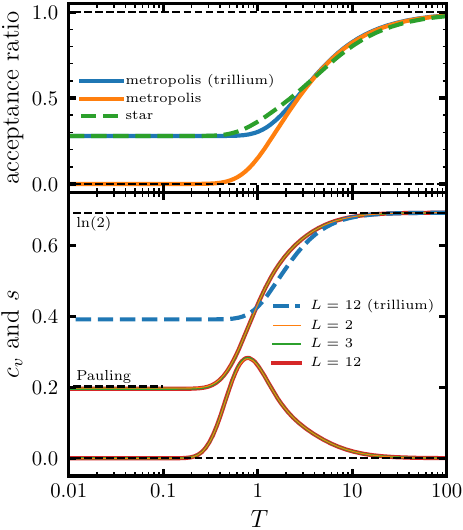}
    \caption{Top panel: acceptance ratio as a function of the temperature for the tetratrillium lattice, using single-spin Metropolis and star updates (orange and green, respectively), compared with the single-spin update on the trillium lattice (blue). Bottom panel: specific heat $c_v$ and entropy $s$ per site as a function of the temperature for the tetratrillium lattice for $L=2$, 3, and 12. The Pauling estimate is shown by a dashed black line. The blue dashed curve corresponds to the entropy in the trillium lattice.}
    \label{fig:cvtetra}
\end{figure}

Next, we discuss the elementary spin moves that carry the system from one ground state to another. Since the ground states in the trillium and tetratrillium cases are in one-to-one correspondence, it is sufficient to discuss only the trillium system; however, we also briefly comment on the generalization to the tetratrillium case below. As mentioned above, the ground states of the trillium lattice consist of triangles in either 2-up-1-down or 1-up-2-down configurations. As a result, situations like the one illustrated in Fig.~\ref{fig:fliploop}(a) (previously discussed in Ref.~\cite{Radpath10}) appear naturally. There, a spin at the corner of three triangles can be flipped without energy cost since it always has one up and one down neighbor in each triangle. Repeating such moves, large numbers of ground state configurations can be visited, resulting in a genuinely disordered system with $\langle S_i \rangle = 0$ for all spins even at $T=0$. The efficiency of these constant-energy single spin moves in exploring the Hilbert space is also indicated in Fig.~\ref{fig:cvtetra} (top panel), showing the acceptance ratio in a Monte-Carlo simulation that makes use only of single spin flips. As shown, the acceptance ratio remains finite for $T\rightarrow0$ in the trillium case (blue curve).

However, these constant-energy single-spin flips do not span the entire ground-state manifold but split up the Hilbert space into disconnected sectors. Specifically, certain ground states are completely isolated in the sense that they do not allow for any constant-energy single-spin moves. This becomes evident already in the $L=1$ lattice, where all six ground states are isolated. By periodically repeating these states, one can construct larger isolated ground states. This implies that for any $L>1$, there are at least six isolated ground states. We can calculate the exact number for $L=2$ and find 42 isolated ground states if only single spin flips are considered. On the other hand, the remaining 314832 states can be reached through single spin flips and compose a connected sub-manifold. 

A natural question is whether these isolated ground states together with the large connected Hilbert space sector are due to a hidden topological structure similar to, e.g., dimer coverings on non-bipartite lattices, which can be classified according to $\mathds{Z}_2$ topological invariants~\cite{Moessner03, Krauth03, Rehn17, Fancelli25}. If that were the case in the trillium system, only non-local updates spanning over the periodic boundaries would connect the aforementioned sectors of isolated and flippable states. On the other hand, local moves involving only $O(1)$ spins could never carry the system between different sectors. However, we identified a counter-example: For systems with $L\geq 5$ constructed by repeating one of the six $L=1$ ground states, a cluster move involving nine spins is possible. This nine-site cluster is formed by two pentagons and one triangle that share the same bond, as presented in Ref.~\cite{Gonzalez24}. This indicates that either the Hilbert space sectors from single spin flips do not follow a topological classification or such a classification cannot be deduced from the small $L=1$ and $L=2$ clusters but requires considering larger systems with $L\geq3$. 

Furthermore, it is also not clear whether the number of isolated states grows extensively or subextensively in system size. However, there is an argument in favor of the latter; the residual entropy from cMC obtained in Ref.~\cite{Radpath10} agrees well with the Pauling estimate, hinting that these states do not contribute to the entropy. A definitive way to confirm this would require calculating the corrections to Pauling entropy using numerical linked cluster expansions, as it was done for the pyrochlore lattice~\cite{Singh12}.

Returning to the tetratrillium lattice, the constant-energy spin flips are no longer possible due to the decorating sites forming tetrahedra, see Fig.~\ref{fig:fliploop}(b). Therefore, the acceptance ratio goes to zero for single-spin (Metropolis) updates, as shown in the top panel of Fig.~\ref{fig:cvtetra} (orange curve). However, it is sufficient to add the three neighboring decorating sites to have a four-spin flippable \textit{star} [shown in red in Fig.~\ref{fig:fliploop}(b)]. Implementing the star flips is straightforward and maintains a finite acceptance ratio at zero temperature, enabling the system to explore the ground-state manifold (see top panel of Fig.~\ref{fig:cvtetra}). Since the ground-state configurations can be mapped one-to-one with those of the trillium lattice, here as well, there are isolated ground-state configurations separated from the remaining manifold by more than one star flip.

\subsubsection{Properties of monomers}

Here, we discuss properties of monomers in the tetratrillium lattice which correspond to tetrahedra with $T_{i,\boxtimes}\neq 0$. We focus on the energetically lowest defect tetrahedra in 1-up-3-down or 3-down-1-up configurations where $T_{i,\boxtimes}=\pm2$. An elementary process that creates two such monomers is the spin flip process $S_i^+ S_j^- + S_i^- S_j^+$ where $i$ and $j$ are nearest neighbor sites on {\it different} trillium lattices. Since such terms correspond to transverse couplings $S_i^x S_j^x+S_i^y S_j^y$, their discussion is relevant for the properties of quantum models that arise upon perturbing the Ising model. As illustrated in Fig.~\ref{fig:fliploop}(c), when applied to a state with all tetrahedra in 2-up-2-down configurations (and assuming that the spin flip operators do not annihilate the state), the monomers are created in two neighboring tetrahedra. A similar spin flip process, where $i$ and $j$ are nearest neighbor sites that both reside on the {\it connected} trillium lattice, creates four monomers and is therefore associated with a larger energy cost.  

Focusing only on the energetically lowest spin flip process that creates two monomers and repeatedly applying these spin flips on different bonds, the monomers can be moved and separated. Specifically, moving two monomers in a loop by bringing them back to each other and annihilating them results in a spin flip process within the ground state manifold. An example of such a process where the loop spans over periodic boundaries is shown in Fig.~\ref{fig:fliploop}(d) (top), where the bonds are flipped in the depicted order of increasing bond thicknesses. At this point, it is instructive to compare the motion of monomers across periodic boundaries in the tetratrillium and pyrochlore lattices. Crucially, the nearest neighbor Ising model on the pyrochlore lattice, a realization of spin ice, can be mapped onto an effective electrodynamics theory~\cite{Krauth03, Hermele04, Benton12}. By identifying the Ising component of the spins with an emergent electric field, the monomer motion through the entire system corresponds to adding an electric field line that winds across periodic boundaries. Importantly, this non-local process carries the system into another topological Hilbert space sector that cannot be accessed with only local spin moves while staying in the spin ice manifold~\cite{Pace21}.

Concerning this last property, the tetratrillium lattice behaves in a crucially different way. The process in Fig.~\ref{fig:fliploop}(d) (top) can, in fact, also be realized by a sequence of local star updates while remaining in monomer-free states, see Fig.~\ref{fig:fliploop}(d) (bottom). Therefore, in the tetratrillium case, the monomer motion across periodic boundaries does not create a new Hilbert space sector. In agreement with the above investigation of disconnected sectors on small spin clusters, this indicates again that the Ising tetratrillium model has no obvious topological structure. Stated differently, monomers in the form of defect tetrahedra do not correspond to the gauge charges of an effective U(1) or $\mathds{Z}_2$ gauge theory. Note, however, that this does not exclude the possibility of a U(1) or $\mathds{Z}_2$ quantum spin liquid in a perturbed Ising tetratrillium model but shows that a possible gauge structure must have a more complicated origin arising as an emergent property. Answering this question, however, goes beyond the scope of this work.

\subsubsection{Thermodynamic behavior}

Returning to the numerical investigation of the tetratrillium Ising model, we use combined Metropolis updates and star flips to perform cMC simulations and calculate the temperature dependence of the entropy and heat capacity per site for different linear system sizes $L$. The results are shown in the bottom panel of Fig.~\ref{fig:cvtetra}. The first interesting feature is that the residual entropy at $T=0$ agrees well with the Pauling estimate (shown by the black dashed line) and the fact that it is half of $s(T=0)$ for the trillium lattice (dashed blue line). Another interesting feature is that the results are already converged for small lattices. We observe a small difference between $L=2$ and $L=3$, and then the results remain practically unchanged up to $L=12$. This means not only that the specific heat does not exhibit a phase transition at finite temperatures, but also that the correlations decay fast with distance, as expected from our large-$N$ analysis. Altogether, these results confirm the presence of a classical spin liquid in the Ising model on the tetratrillium lattice.

\begin{figure}[t!]
    \centering
    \includegraphics[width=0.9\linewidth]{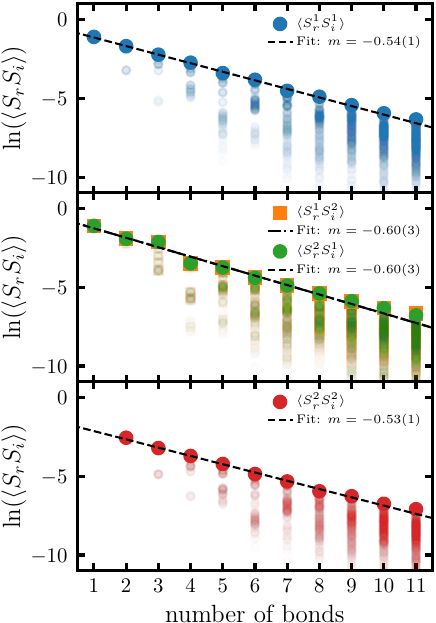}
    \caption{Spin-spin correlation decay as a function of the distance (measured in number of bonds between sites) for $T=0.01$. The correlations are taken between sites in the same and different sublattices, indicated by the superscripts in the legends, where $S_r$ is the reference site. For each set and for each distance, the highest value of the correlation is shown in larger and solid symbols. This data is used to perform a linear fit with slope $m$.}
    \label{fig:corrs}
\end{figure}

\subsubsection{Spin correlations}

Finally, we confirm that this classical spin liquid is the same as that we previously predicted with the large-$N$ theory by calculating the spin correlations at low temperatures. We show in the first row of Fig.~\ref{fig:ssfall} the results of the spin structure factor at $T=0.01$, which are in complete agreement with the results discussed previously in the large-$N$ limit in the third row. As the spin liquid predictions from the large-$N$ limit seem to hold, the correlations should decay exponentially due to the gapped nature of the spectrum. To confirm this, we show in Fig.~\ref{fig:corrs} the resulting correlations at $T=0.01$ as a function of the distance between spins measured via the number of bonds (the shortest number of bonds to reach a given spin). We perform calculations using reference sites in the two sublattices, $S_r^1$ and $S_r^2$, and divide our results into three different types of correlations. In the top panel, we show correlations corresponding to the connected trillium lattice. In the middle one, we show correlations between sublattices, and in the last one, correlations in the disconnected trillium lattice (where there is no data at distance 1). For all distances, we take the largest correlations to perform the linear fit, while the remaining data is shown with transparency. In all cases, it becomes evident that the correlations decay exponentially as manifested by the linear fits in the log plot. Even though the slopes of the fits, $m$, seem to be different in the middle panel, they are within error bars when changing the range of the fit.

\subsection{The Heisenberg case $(N=3)$}

We now move on to the classical Heisenberg model, which we again treat using cMC calculations. It is important to investigate this model separately since, in some cases, the results from the large-$N$ limit do not hold in the more realistic Heisenberg ($N=3$) limit. Two examples of this are the trillium and kagome lattices. For the former, even though the large-$N$ theory predicts a spiral liquid, the ground state in the Heisenberg limit is unique and there is a finite-temperature phase transition at $T_c=0.21~J$ into a magnetically ordered state (but no order-by-disorder selection or subextensive degeneracy)~\cite{Isakov08}. Nevertheless, the spin structure factors of the Heisenberg and large-$N$ cases are quantitatively similar at finite temperatures above the transition. For the kagome lattice, the large-$N$ theory predicts an extensively degenerate spin liquid~\cite{Garanin99}. However, the Heisenberg limit exhibits an order-by-disorder selection of special coplanar states~\cite{Chalker92, Zhitomirsky08, Chern13}.

\begin{figure}[t!]
    \centering
    \includegraphics[width=0.9\linewidth]{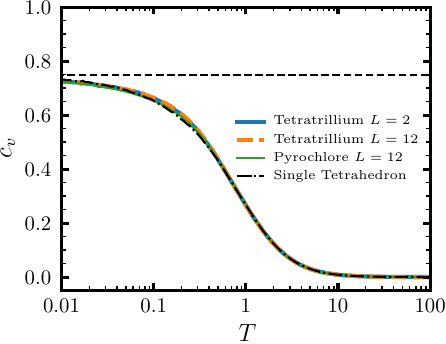}
    \caption{Specific heat per site of the classical Heisenberg model on the tetratrillium lattice ($L=2$ in blue and $L=12$ in dashed orange), the pyrochlore lattice ($L=12$ in green), and a single tetrahedron multiplied by 2 (dot-dashed black). The horizontal dashed black line indicates $c_v = 0.75$.}
    \label{fig:cvtetraheis}
\end{figure}

Our specific heat results for the tetratrillium lattice are shown in Fig.~\ref{fig:cvtetraheis}, where systems with $L=2$ and 12 are shown in blue and orange, respectively. Again, as in the case of the Ising model, we observe very little change with system size and the absence of a finite-temperature phase transition, indicating a disordered ground state. Furthermore, there are only slight changes when compared to the thermodynamics of a single classical tetrahedron (multiplied by 2 to match the number of bonds per spin, shown in a dot-dashed black line). These observations were already made in Ref.~\cite{Gonzalez24}, where it was argued that the negligible change with system size could be attributed to the short-range correlations. However, in Fig.~\ref{fig:cvtetraheis} we also show the specific heat calculations for the pyrochlore Heisenberg antiferromagnet (green line). Surprisingly, and despite having algebraically decaying correlations, the specific heat is very similar to that of the tetratrillium lattice (albeit exhibiting some clear differences at low temperatures). 

\begin{figure}[t!]
    \centering
    \includegraphics[width=0.9\linewidth]{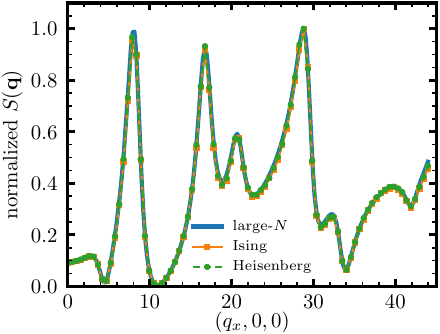}
    \caption{Spin structure factor at $T=0.01$ along the $(q_x, 0, 0)$ direction for the large-$N$ (blue line), Ising (orange squares and line), and Heisenberg cases (green circles and dashed line).}
    \label{fig:ssfcut}
\end{figure}

Altogether, the specific heat in both the tetratrillium and pyrochlore lattices shows an excellent agreement with the specific heat of a single tetrahedron above $T\sim J$. This is expected since both lattices are composed of corner-sharing tetrahedra, and the inter-tetrahedra correlations are irrelevant at large temperatures. The other common feature in all systems is that $c_v(T\to0)=0.75$. As explained in Refs.~\cite{Moessner98, Moessner98b, Gonzalez24}, this can be attributed to the absence of quartic modes associated with magnetic order. Despite the similarities in the specific heat, the pyrochlore and tetratrillium lattices are fundamentally different, mainly because the former is bipartite in terms of tetrahedra while the latter is not. This is the reason for the gapped nature of the tetratrillium lattice in the large-$N$ limit, leading to exponentially decaying correlations (as shown in the Ising case). 

The spin structure factor calculations at $T=0.01$ are shown in the second row of Fig.~\ref{fig:ssfall}, where it becomes evident that they match the previously shown results on the Ising and large-$N$ limiting cases. This means that the structure of correlations in the Heisenberg case is the same as the one corresponding to the classical fragile spin liquid predicted by the large-$N$ theory. To further emphasize the quantitatively excellent agreement in the different $N$ cases, we plot in Fig.~\ref{fig:ssfcut} the spin structure factors along the $(k_x, 0, 0)$ direction for the large-$N$, Ising, and Heisenberg cases.

\subsection{Quantum $S=1/2$ Heisenberg model}

As our final model, we now investigate the effects of quantum fluctuations within PMFRG by transforming the 3-component spin vectors into quantum-mechanical operators. In contrast with the results presented in Ref.~\cite{Gonzalez24} for the $S=1$ system K$_2$Ni$_2$(SO$_4$)$_3$, here we treat the $S=1/2$ case as the limit where quantum fluctuations are strongest. Another difference is that, so far, only the PFFRG method has been applied to this system, which is a $T=0$ method that works at finite cutoff frequency $\Lambda$ and recovers the ground state only in the $\Lambda \to 0$ limit. While $\Lambda$ acts similarly to an effective temperature, the detection of symmetry breaking relies on the search for anomalies in the flow of the susceptibility, which works well to distinguish clearly ordered phases but becomes imprecise in the vicinity of spin liquid phases. In the PMFRG framework, the identification of magnetic long-range order is facilitated by the possibility of performing finite-size scaling analyses at temperatures $T>0$.

As mentioned in Sec.~\ref{sec:modandmet}, the lattices we consider are technically infinite due to the conserved translation symmetries. However, a certain finite size cutoff is introduced by $N_\textrm{len}$, which determines the radius up to which correlations are considered (outside the sphere, all correlations are set to zero). Here we consider $N_\textrm{len}$ from 1.5 to 5.0 (from 86 to 2804 inequivalent bonds), where the cubic unit cell has the side length $a=1$.

\begin{figure}[t!]
    \centering
    \includegraphics[width=0.9\linewidth]{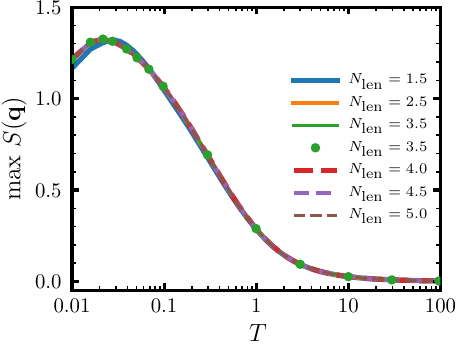}
    \caption{Maximum of the spin structure factor $S(\mathbf{q})$ as a function of the temperature for different lattice sizes $N_\textrm{len}$, obtained with PMFRG. The dots indicate the temperatures where $q^\textrm{max}$ has been found by calculating the complete spin structure factor for $N_\textrm{len}=3.5$. The lines correspond to maximization within the selected $q^\textrm{max}$ (see main text).}
    \label{fig:ssfmax}
\end{figure}

Our approach for detecting a possible finite-temperature phase transition consists of identifying peaks in the spin structure factor $S(\mathbf{q})$, which appear at wave vectors $q^\textrm{max}$, as a function of the temperature for different lattice sizes. Taking into account that $S(\mathbf{q})$ is non-periodic in momentum space due to the incommensurate positions of the spins within the unit cell, it is computationally expensive to find $q^\textrm{max}$ for every temperature. Therefore, we calculate $q^\textrm{max}$ only for some selected temperatures (shown by green circles in Fig.~\ref{fig:ssfmax}), by considering the $q$-discretization of an $L=40$ lattice and using results for $N_\textrm{len}=3.5$. We find that only two wave-vectors $q^\textrm{max}$ compete above $T=0.05$, and others appear for smaller temperatures. Then, we maximize $S(\mathbf{q})$ within the subspace of these $\{q^\textrm{max}\}$ for all system sizes. The result is shown by the lines in Fig.~\ref{fig:ssfmax}, where it becomes clear that the results converge above $N_\textrm{len}=2.5$, with only the $N_\textrm{len}=1.5$ curve differing from the others. This can be interpreted as a decay of the correlations to negligible values for distances larger than two and a half unit cells, which implies that there is no tendency towards long-range order. The anomalous behavior below $T=0.05$ suggests that the results of the RG flow cannot be trusted in this low temperature regime. This is also reflected in the presence of negative (i.e., unphysical) values of the spin structure factor appearing near $\mathbf{q} = \mathbf{0}$. 

Finally, in the bottom row of Fig.~\ref{fig:ssfall} we show the results of the spin structure factor in the same three planes discussed for the other models. As explained before, the results at very low temperatures cannot be trusted because unphysical negative values start appearing in the spin structure factor. Therefore, we show results at $T=0.1$, which is still in the regime where quantum fluctuations are expected to dominate over thermal fluctuations. Our results confirm the existence of a disordered state with a broadened spin structure factor and no ordering tendency (sharp peaks). Notably, some patterns can be clearly identified as broadened features from the classical cases, but some features appear purely in the quantum case. Furthermore, there are clear similarities when comparing these results to the experimental measurements on the $S=1$ K$_2$Ni$_2$(SO$_4$)$_3$~\cite{Gonzalez24}.

\section{Conclusion}
\label{sec:conc}

In this article, we have studied the spin liquids that arise in the tetratrillium lattice. This lattice, relevant to the Langbeinite compound K$_2$Ni$_2$(SO$_4$)$_3$, consists of a trillium lattice in which each triangle is transformed into a tetrahedron. We first studied the classical limits and found that in the large-$N$, Ising, and Heisenberg cases, the correlations at low temperatures show the same structure, indicating the same type of disordered state. From the large-$N$ limit, we found that the spectrum of the coupling matrix presents a gap between the four flat bands at the bottom and the dispersive bands. According to the classical spin-liquid classification scheme~\cite{Yan24a, Yan24b, Davier23}, this corresponds to a rare example of a fragile topological spin liquid in three dimensions, in which the correlations decay exponentially even at $T=0$. Such phases, contrary to the more common U(1) spin liquids, do not present any pinch-point singularities in the spin structure factor, associated with emergent Gauss laws.

We also studied the quantum $S=1/2$ Heisenberg case and found no ordering signatures in the spin structure factor down to low temperatures. However, the reliable identification of a spin-liquid ground state remains practically out of reach within PMFRG or any other available numerical many-body method, particularly for frustrated three-dimensional lattices. Alternatively, it would be interesting to study the quantum ground state that results from adding perpendicular spin interactions to the Ising case. This would allow one to study the effect of quantum fluctuations over the extensive ground-state manifold perturbatively, as was done in the well-known case of quantum spin ice from classical spin ice. In this article, we have taken first steps in this direction and found that, in contrast to quantum spin ice on the pyrochlore lattice, defect tetrahedra do not represent the gauge charges of an underlying $\mathds{Z}_2$ or U(1) gauge theory, and that the Ising configurations exhibit no obvious topological sectors. This observation implies that the true nature of the quantum ground state remains essentially unresolved. It could host a range of possible disordered phases with emergent gauge structures that are not directly dictated by the underlying lattice geometry of corner-sharing tetrahedra. Exploring this question constitutes a promising direction for future research, with direct relevance to real magnetic materials.

\section*{Acknowledgements}

The authors thank Vincent Noculak, Yannik Schaden, Yasir Iqbal, Jeffrey Rau, and Ludovic Jaubert for fruitful discussions. M. G. G. gratefully acknowledges the use of computing time at the HPC systems Curta of the Freie Universität Berlin and Bonna of the University of Bonn. J. R. acknowledges support from the Deutsche Forschungsgemeinschaft (DFG, German Research Foundation), within Project-ID 277101999 CRC 183 (Project A04).

\bibliography{biblio}

\end{document}